\documentclass{llncs2e/llncs}
\usepackage{makeidx}  
\usepackage{mystyle}

\begin{document}
\frontmatter          
\pagestyle{headings}  
\addtocmark{GD-IMLOP} 
%
%
\mainmatter              
\title{Endoscopic navigation in the absence of CT imaging}
\titlerunning{EndNav_woCT}  
%
\author{Ayushi Sinha\inst{1} 
\and Xingtong Liu\inst{1}  \and Austin Reiter\inst{1}
\and Masaru Ishii\inst{2}\and Gregory~D.~Hager\inst{1} \and Russell~H.~Taylor\inst{1}}
%
\authorrunning{Sinha et al.} 
%

\tocauthor{Ayushi Sinha, Xingtong Liu, Austin Reiter, 
Masaru Ishii, Gregory D. Hager, Russell H. Taylor}
%
\institute{The Johns Hopkins University, Baltimore, USA,\\
\email{sinha@jhu.edu}
\and
Johns Hopkins Medical Institutions, Baltimore, USA \\}

\maketitle              

\begin{abstract}
Clinical examinations that involve endoscopic exploration of the nasal cavity and sinuses often do not have a reference image to provide structural context to the clinician. 
In this paper, we present a system for navigation during clinical endoscopic exploration in the absence of computed tomography (CT) scans by making use of shape statistics from past CT scans. Using a deformable registration algorithm along with dense reconstructions from video, we show that we are able to achieve submillimeter registrations in in-vivo clinical data and are able to assign confidence to these registrations using confidence criteria established using simulated data. 


\end{abstract}
\section{Introduction}
\label{sec:intro}
Endoscopic explorations of the nasal cavity and sinuses are generally not accompanied by a reference computed tomography (CT) image since CT image acquisition exposes patients to high doses of ionizing radiation and is, therefore, avoided unless necessary. Clinicians performing the exploration must rely entirely on the endoscopic camera for visualization and, therefore, must cope with restricted field of view. In order to reduce reliance on experience or memory and to provide additional context information, we have developed a system that enables navigation without the need for accompanying patient CT or other similar imaging and associates a confidence measure to the navigation being provided. Further, our system does not introduce any additional devices than those already used in clinical endoscopic exploration. Therefore, the clinician is not responsible for anything in addition to the endoscope.

Most navigation systems that have been developed are intended for surgical use~\cite{Mirota11,Azagury12}. 
For surgical navigation, there is almost always access to preoperative CT scans, which have high contrast between air, bone, and soft tissue. This allows surgeons to better understand their location, the proximity to surrounding bones and soft tissue, and the thickness of surrounding bones, enabling them to make more informed decisions during surgery and prevent harm to critical structures nearby, like the brain, eyes, optic nerves, carotid arteries, etc.
%
%

The main difference between these previous methods and the method presented here is the absence of patient specific CT scans. In order to make up for this absence, we utilize past CT scans to build statistical shape models of relevant structures. Statistically derived shapes are then deformably registered to dense reconstructions of anatomy visible in endoscopic video, and statistical confidence measures are automatically assigned to the registrations. The registration accomplishes two tasks simultaneously. First, it aligns the endoscopic video to the statistically derived shape, giving the clinician more information about where surrounding structures 
may be. Second, it deforms the statistically derived shape to fit the structure obtained from video and, in effect, \emph{estimates the patient CT}. The confidence measure further informs the clinician on when and how much the navigation system can be trusted, and also allows the navigation system to attempt to improve itself if its current registration estimate has low confidence.
We perform two experiments to evaluate our 
framework. First, we 
establish that our framework can compute submillimeter registrations and reliably assign confidence to the registrations using simulated data. 
Second, we evaluate our framework on in-vivo clinical data, and use the confidence criteria 
to assign confidence to the registrations.


%
\section{Method}
\label{sec:method}
%
%
To build statistical shape models (SSMs), 
we automatically segment $53$ publicly available head CTs~\cite{Beichel15, Bosch15, Clark13, Fedorov16} by transferring 3D meshes extracted from manually created labels in a template CT image to the $53$ CTs using deformation fields produced by an intensity-based CT-CT registration algorithm~\cite{Avants11}. With some improvement to these initial segmentations using the method described in \cite{Sinha17}, we obtain reliably segmented structures in all CTs along with reliable correspondences. These correspondences allow us to build SSMs of the segmented structures using established methods like principal component analysis (PCA)~\cite{Cootes95}:
\begin{equation}\label{eq:modes}
	\Sigmassm = \frac{1}{\n_s}\sum_{j=1}^{\n_s} (\V_j - \bar{\V})\trans(\V_j-\bar{\V})	= [\m_1 \ldots \m_{\n_s}]
				\begin{bmatrix}
       					\lambda_1 	&  		& 	\\
       					 		& \ddots 	&	\\
      					 		& 		& \lambda_{\n_s}
     				\end{bmatrix}
	[\m_1 \ldots \m_{\n_s}]\trans, 
\end{equation}
where $\V_j$ is the stacked vector of vertices, $\V = [\vv_1 \  \vv_2 \ldots \vv_{\n_v}]\trans$, for the $j$th mesh, $\bar{\V}$ is the mean shape computed by averaging the $\n_v$ corresponding vertices over $\n_s$ shapes, $\bar{\V} = \frac{1}{n_s}\sum_{j=1}^{n_s} \V_j$, and $\Sigmassm$ is the shape covariance matrix. An eigen decomposition of $\Sigmassm$ produces the principal modes of variation, $\m$, and the mode weights, $\lambda$, which represent the amount of variation 
along the corresponding $\m$ (Eq.~\ref{eq:modes}). PCA enables any new shape, $\V^*$, that is in correspondence 
with the shapes used to build the SSM, to be estimated using $\bar{\V}$, $\m$ and $\lambda$: 
$\tilde{\V}^* = \bar{\V} + \sum_{j=1}^{\n_m} s_j\w_j,$
where $\tilde{\V}^*$ is the estimated 
$\V^*$, $1 \leq \n_m < \n_s$ is some specified number of modes, $\w_j = \sqrt{\lambda_j}\m_j$ are the weighted modes of variation, and $s_j$ are the shape parameters in units of standard deviation (SD) which can be obtained by projecting the mean subtracted $\V^*$ onto the weighted modes.

%
These shape parameters, $\s = \{s_j\}$, can be incorporated into probabilistic models of registration to enable optimization over $\s$ in addition to other registration parameters~\cite{Sinha18}. In particular, we evaluate the deformable extension of the generalized iterative most likely oriented point (G-IMLOP) algorithm, an iterative rigid registration algorithm~\cite{Billings15b}. 
The generalized \emph{deformable} iterative most likely oriented point (GD-IMLOP) algorithm extends G-IMLOP, which incorporates an anisotropic Gaussian noise model and an anisotropic Kent noise model to account for measurement errors in position and orientation, respectively~\cite{Billings15b}. Assuming both position and orientation errors are zero-mean, independent and identically distributed, the match likelihood function for each oriented point, $\x$, transformed by a current similarity transform, $[a,\R,\tv]$, is defined as~\cite{Billings15b}:
\begin{equation}
\begin{split}
	&\fmatch(\x; \y ,\Sigmax, \Sigmay, \kappa, \beta, \hat{\gammab}_1, \hat{\gammab}_2, a, \R, \tv) = \frac{1}{\sqrt{(2\pi)^3 |\Sigmab|}\cdot c(\kappa,\beta)}\\
	&\cdot e^{-\frac{1}{2}(\y_\p-a\R\x_\p-\tv)\trans\Sigmab\inv (\y_\p-a\R\x_\p-\tv)-\kappa\hat{\y}_\n\trans\R\hat{\x}_\n + \beta\left( \left( \hat{\gammab}_1\trans\R\hat{\x}_\n \right)^2-\left( \hat{\gammab}_2\trans\R\hat{\x}_\n \right)^2 \right)}.
\label{eq:f_gdimlop}
\end{split}
\end{equation}
This 
function finds the $\y = (\y_\p, \hat{\y}_\n)$ that maximizes the likelihood of a match with $\x = (\x_\p, \hat{\x}_\n)$. $\Sigmab = \R\Sigmax\R\trans + \Sigmay$, where $\Sigmax$ and $\Sigmay$ are the covariance matrices representing the measurement noise associated with $\x$ and $\y$, $\kappa= \frac{1}{\sigma^2}$ is the concentration parameter of the orientation noise model, where $\sigma$ is the SD of orientation noise, and $\beta= e\frac{\kappa}{2}$ controls the anisotropy of the orientation noise model along with $\hat{\gammab}_{1}$ and $\hat{\gammab}_{2}$, which are the major and minor axes that define the directions of the elliptical level sets of the Kent distribution on the unit sphere~\cite{Billings15b, Mardia08}. $\hat{\y}_\n$, $\hat{\gammab}_{1}$, $\hat{\gammab}_{2}$ are orthogonal and $e \in [0,1]$ is the eccentricity of the noise model. 

Correspondences are computed by minimizing the negative log likelihood of $\fmatch$~\cite{Sinha18}. The main difference in the correspondence phases of G-IMLOP and GD-IMLOP is that GD-IMLOP computes matched points on the current \emph{deformed} shape. 
Outlier rejection is performed after each correspondence phase. Under the assumption of generalized Gaussian noise, the square Mahalanobis distance is approximately distributed as a chi-square distribution with $3$ degrees of freedom (DOF)~\cite{Billings15b}. Therefore, a match is labeled an outlier if this distance exceeds the value of a chi-square inverse cumulative density function (CDF) with $3$ DOF at some probability $p$. That is, if for any corresponding $\x$ and $\y$,
$(\y_\p-a\R\x_\p-\tv)\trans\Sigmab\inv (\y_\p-a\R\x_\p-\tv) > \chisqinv(p, 3),$
then that match is an outlier. Here, we set $p=0.95$. Matches that are not rejected as outliers using this test, 
are evaluated for orientation consistency. Here, a match is an outlier if $\hat{\y}_\n\trans\R\hat{\x}_\n < \cos{(\thetathresh)}$, where $\thetathresh = 3\sigmacirc$ and $\sigmacirc$ is the circular SD computed using the mean angular error between all correspondences. 

Matches that pass these two tests are inliers and a registration between these points is computed by minimizing the following cost function with respect to the transformation and shape parameters~\cite{Sinha18}:
\small
\begin{equation}
\begin{split}
&\T = \argmin_{[a,\R,\tv],\s}\Bigg(\frac{1}{2}\sum_{i=1}^{\ndata}\Big((\Tssm(\ypi)-a\R\xpi-\tv)\trans\Sigmab\inv (\Tssm(\ypi)-a\R\xpi-\tv)\Big) + \\
&\sum_{i=1}^{\ndata} \kappa_i (1-\ynitrans\R\xni) - \sum_{i=1}^{\ndata} \beta_i\left( \left( \hat{\gammab}_{1i}\trans\R\trans\hat{\y}_{\n i} \right)^2 - \left( \hat{\gammab}_{2i}\trans\R\trans\hat{\y}_{\n i} \right)^2 \right) + \frac{1}{2}\sum_{j=1}^{n_\m}\norm{s_j}_2^2\Bigg),
\label{eq:C_gdimlop}
\end{split}
\end{equation}
\normalsize
where $\ndata$ is the number of inlying data points, $\x_i$. 
This first term in Eq.~\ref{eq:C_gdimlop} minimizes the Mahalanobis distance between the positional components of the correspondences, $\xpi$ and $\ypi$. $\Tssm(\cdot)$, a term introduced in the registration phase, is a transformation, $\Tssm(\ypi) = \sum_{j=1}^3 \mu_i^{(j)}\Tssm(\vv_i^{(j)})$, 
that deforms the matched points, $\y_i$, based on the current $\s$ deforming the model shape~\cite{Sinha18}. Here, $\Tssm(\vv_i) = \bar{\vv}_i + \sum_{j=1}^{\n_m} s_j\w_j^{(i)}$, and $\mu_i^{(j)}$ are the $3$ barycentric coordinates that describe the position of 
$\y_i$ on a triangle on the model shape~\cite{Sinha18}. The second and third terms minimize the angular error between the orientation components of corresponding points, $\xni$ and $\yni$, while respecting the anisotropy in the orientation noise. The final term minimizes the shape parameters to find the smallest deformation required to modify the model shape to fit the data points, $\x_i$~\cite{Sinha18}. $\s$ is initialized to $0$, meaning the registration begins with the statistically mean shape. 
The objective function (Eq.~\ref{eq:C_gdimlop}) is optimized using a nonlinear constrained quasi-Newton based optimizer, where the constraint is used to ensure that $\s$ are found within $\pm3$ SDs, since this interval explains $99.7\%$ of the variation.

Once the algorithm has converged, a final set of tests is performed to 
assign confidence to the computed registration. For position components, this is similar to the outlier rejection test, except now the sum of the square Mahalanobis distance is compared against the value of a chi-square inverse CDF with $3\ndata$ DOF~\cite{Billings15b}; i.e., confidence in a registration begins to degrade if 
\begin{equation}
\begin{split}
\mathrm{E}_p = \sum_{i=1}^{\ndata}(\ypi-a\R\xpi-\tv)\trans\Sigmab\inv (\ypi-a\R\xpi-\tv) > \chisqinv(p, 3\ndata).
\label{eq:pos_rej}
\end{split}
\end{equation}
If 
a registration is successful according to Eq.~\ref{eq:pos_rej}, it is further tested for orientation consistency using a similar chi-square test by approximating the Kent distribution as a 2D wrapped Gaussian~\cite{Mardia08}. Registration confidence degrades if 
\begin{equation}
\begin{split}
\mathrm{E}_o = \sum_{i=1}^{\ndata}
\begin{bmatrix}
    \cos\inv{(\yni\trans\R\xni)}  \\
    \sin\inv{(\hat{\gammab}_{1_i}\trans\R\trans\hat{\y}_{\n_i})}  \\
    \sin\inv{(\hat{\gammab}_{2_i}\trans\R\trans\hat{\y}_{\n_i})} 
\end{bmatrix}^\trans
\begin{bmatrix}
    \kappa_i 	& 		0 		& 0  \\
    0       	& \kappa_i - 2\beta_i 	& 0 \\
    0       	&		0 		& \kappa_i+2\beta_i
\end{bmatrix}
\begin{bmatrix}
    \cos\inv{(\yni\trans\R\xni)}  \\
    \sin\inv{(\hat{\gammab}_{1_i}\trans\R\trans\hat{\y}_{\n_i})}  \\
    \sin\inv{(\hat{\gammab}_{2_i}\trans\R\trans\hat{\y}_{\n_i})} 
\end{bmatrix} &\\
> \chisqinv(p, 2\ndata),&
\label{eq:ang_rej}
\end{split}
\end{equation}
since $\yni$ must align with $\xni$, but remain orthogonal to $\hat{\gammab}_{1_i}$ and $\hat{\gammab}_{2_i}$. $p$ is set to $0.95$ for \textbf{very confident} success classification. 
As $p$ increases, the confidence in success classification decreases while that in failure classification increases.
\section{Experimental results and discussion}
\label{sec:exp}
Two experiments are conducted to evaluate this system: one using simulated data where ground truth is known, and one using in-vivo clinical data where ground truth is not known. Registrations are computed using $\n_m \in \{0, 10, 20, 30, 40, 50\}$ modes. 
At $0$ modes, this algorithm is essentially G-IMLOP with an additional scale component in the optimization. 

\begin{figure}[b]
\begin{center} 
  \includegraphics[width=0.32\linewidth]{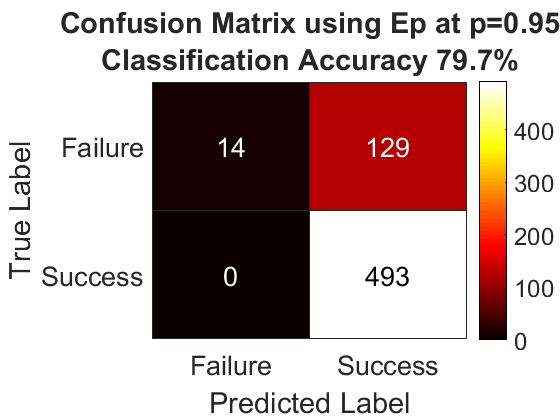} 
  \includegraphics[width=0.32\linewidth]{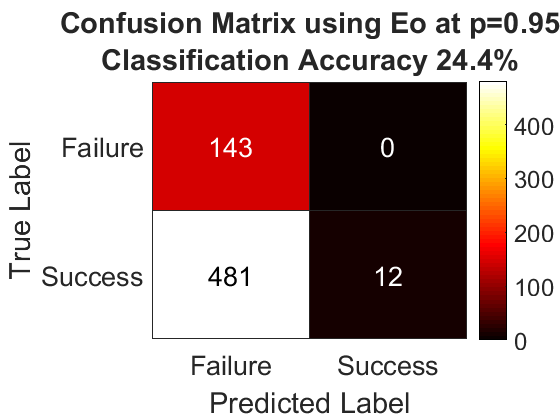} 
  \includegraphics[width=0.32\linewidth]{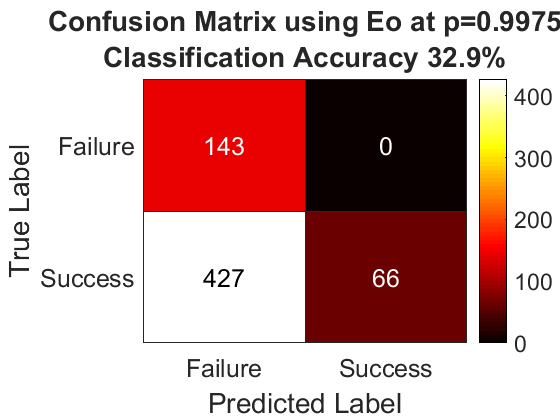} 
  \caption{Left: Using only $\mathrm{E}_p$, all successful registration pass the chi-square inverse test at $p=0.95$. However, many failed registrations also pass this test. Using $p=0.9975$ produces the same result. Middle: On the other hand, using only $\mathrm{E}_o$, no failed registrations pass the chi-square inverse test at $p=0.95$, but very few successful registrations pass the test. Right: Using 
$p=0.9975$, more successful registrations pass the test.}
  \label{fig:cm}
\end{center}
\end{figure}

\subsection{Experiment 1: Simulation}
\label{sec:exp_sim}
In this experiment, we performed a leave-one-out evaluation using shape models of the right nasal cavity extracted from $53$ CTs. $3000$ points were sampled from the section of the left out mesh that would be visible to an endoscope 
inserted into the cavity. Anisotropic noise with SD $0.5\times0.5\times0.75$\,mm$^3$ and $10^\circ$ with $e=0.5$ was added to the position and orientation components of the 
points, respectively, since this produced realistic point clouds compared to in-vivo data with higher uncertainty in the z-direction. 
A rotation, translation and scale are applied to these 
points in the intervals $[0,10]$\,mm, $[0,10]^\circ$ and $[0.95, 1.05]$, respectively. $2$ offsets are sampled 
for each left out shape. GD-IMLOP makes slightly more generous noise assumptions with SDs $1\times1\times2$\,mm$^3$ and $30^\circ$ $(e=0.5)$ for position and orientation noise, respectively, and restricts scale optimization to within $[0.9, 1.1]$. A registration is considered successful if the total registration error (tRE), computed using the Hausdorff distance (HD) between the left-out shape and the estimated shape transformed to the frame of the registered points, is below $1$\,mm. The success or failure of the registrations is compared to the outcome predicted by the algorithm. 
Further, the HD between the left-out and estimated shapes in the same frame 
is used to evaluate errors in reconstruction.

\begin{figure}[b]
\begin{center} 
  \includegraphics[width=0.33\linewidth]{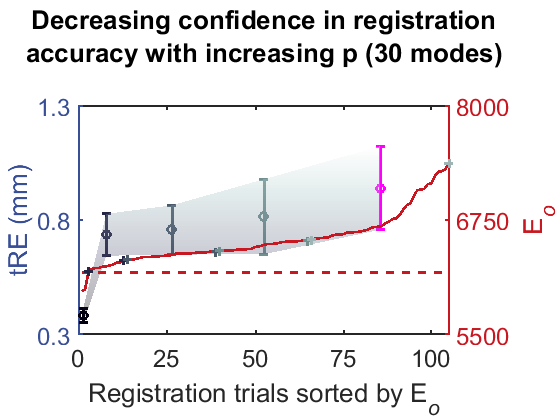} 
  \includegraphics[width=0.33\linewidth]{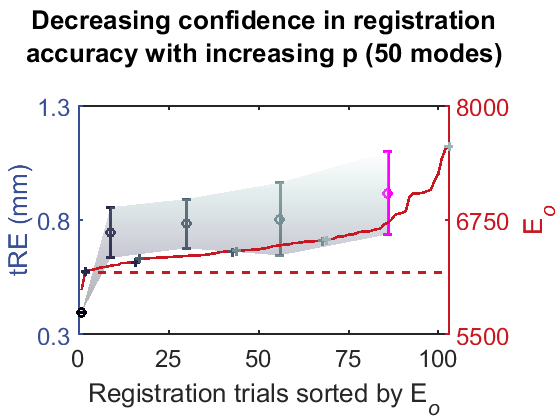} 
  \includegraphics[width=0.323\linewidth]{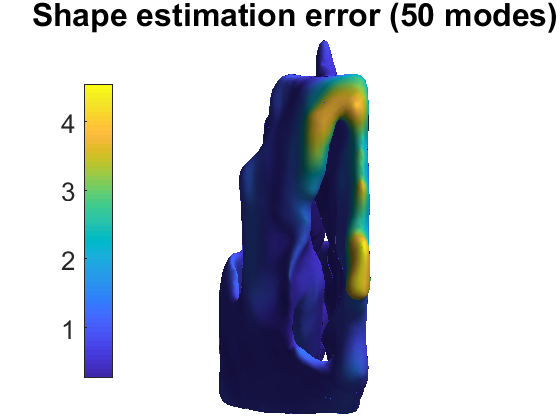} 
  \caption{Left and middle: Mean tRE and standard deviation increase as $\mathrm{E}_o$ increases. The dotted red line corresponds to $\chisqinv(0.95, 2\ndata)$, below which registrations are classified \textbf{very confidently} as successful. Beyond this threshold, confidence gradually degrades. The pink bar indicates that none of these registrations passed the $\mathrm{E}_p$ test. Right: Average error at each vertex computed over all left-out trials using $50$ modes.}
  \label{fig:deteriorate}
\end{center}
\end{figure}

Results over all modes, using $p=0.95$, show that $\mathrm{E}_p$ is less strict than $\mathrm{E}_o$ (Fig.~\ref{fig:cm}), meaning that although $\mathrm{E}_p$ identifies all successful registrations correctly, it also allows many unsuccessful registrations to be labeled successful. $\mathrm{E}_o$, on the other hand, correctly classifies fewer successful registrations, but does not label any failed registrations as successful. Therefore, registrations 
with $\mathrm{E}_p < \chisqinv(0.95, 3\ndata)$ and $\mathrm{E}_o < \chisqinv(0.95, 2\ndata)$ can be \textbf{very confidently} classified as successful. The average tRE produced by registrations in this category over all modes was $0.34$ $(\pm 0.03)$\,mm. 
At $p=0.9975$, 
more successful registrations were correctly identified (Fig.~\ref{fig:cm}, right). These registrations can be \textbf{confidently} classified as successful with mean tRE increasing to $0.62$ $(\pm 0.03)$\,mm. Errors in correct classification creep in with $p=0.9999$, where $3$ out of $124$ registrations are incorrectly labeled successful. These registrations can be \textbf{somewhat confidently} classified as successful with mean tRE increasing slightly to $0.78$ $(\pm 0.04)$\,mm. Increasing $p$ to $0.999999$ further decreases classification accuracy. $10$ out of $121$ registrations in this category are incorrectly classified as successful with mean tRE increasing to $0.8$ $(\pm 0.05)$\,mm. These registrations can, therefore, be classified as successful with \textbf{low confidence}. The mean tRE for the remaining registrations increases to over $1$\,mm at $1.31$ $(\pm 0.85)$\,mm, with no registration passing the $\mathrm{E}_p$ threshold except for registrations using $0$ modes. Of these, however, $0$ are correctly classified as successful. Therefore, although about half of all registrations 
in this category are successful, there can be \textbf{no confidence} in their correct classification. 
Fig.~\ref{fig:deteriorate} (left and middle) shows the distribution of tREs in these categories for registrations using $30$ and $50$ modes. 

GD-IMLOP can, therefore, compute successful registrations between a statistically mean right nasal cavity mesh and points sampled only from 
part of the left-out 
meshes, and reliably assign confidence to these registrations. Further, GD-IMLOP can accurately estimate the region 
of the nasal cavity where 
points are sampled from, 
while errors gradually deteriorate away from this region, 
e.g., towards the front of the septum 
since points are not sampled from this region (Fig.~\ref{fig:deteriorate}, right). Overall, the mean shape estimation error was $0.77$\,mm.

\begin{figure}[b]
\begin{center} 
  \includegraphics[width=0.21\linewidth]{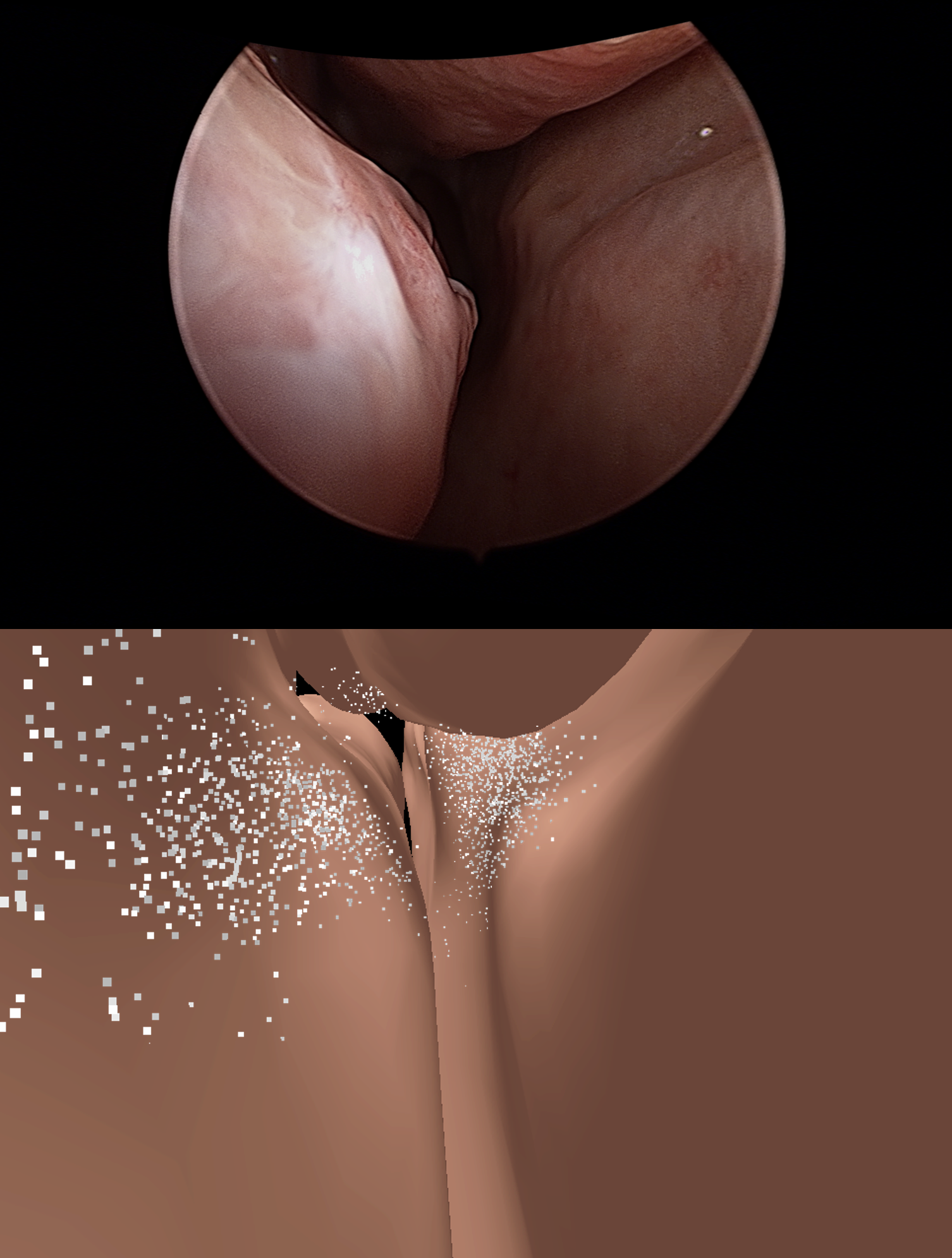} 
  \includegraphics[width=0.37\linewidth]{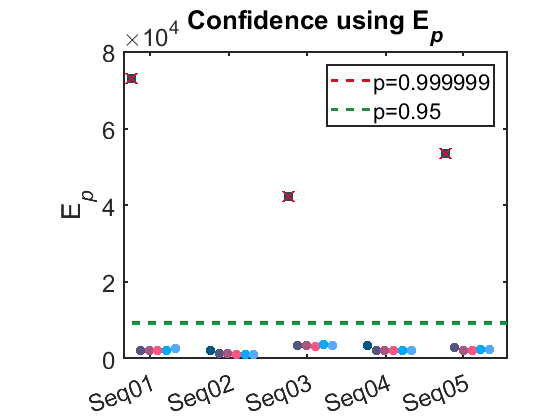} 
  \includegraphics[width=0.37\linewidth]{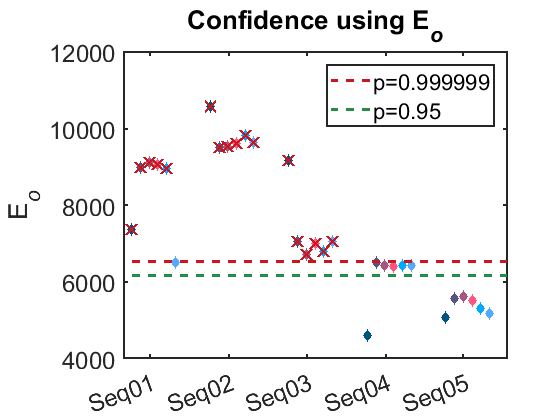} 
  \caption{Left: Visualization of the final registration and reconstruction for Seq01 using 50 modes. Middle and right: $\mathrm{E}_p$ and $\mathrm{E}_o$ for all registrations, respectively, plotted for each sequence. Per sequence, from left to right, the plot points indicate scores achieved using $0$-$50$ modes at increments of $10$. Crossed out plot points indicate rejected registrations.}
  \label{fig:clinic}
\end{center}
\end{figure}

\subsection{Experiment 2: In-vivo}
\label{sec:exp_clinic}
For the in-vivo experiment, we collected 
anonymized 
endoscopic videos of the nasal cavity 
from consenting patients under an IRB approved study. Dense point clouds 
were produced from single frames of these videos using a modified version 
of the learning-based photometric reconstruction technique~\cite{Reiter16} that uses registered structure from motion (\SfM) points
to train a neural network to predict dense depth maps. 
Point clouds from different nearby frames in a sequence were aligned using the relative camera motion from \SfM. 
Small misalignments due to errors in depth estimation were corrected using 
G-IMLOP with scale 
to produce a dense reconstruction spanning a large area of the nasal passage. 
GD-IMLOP is executed with $3000$ points sampled from this dense reconstruction assuming noise with SDs $1\times1\times2$\,mm$^3$ and $30^\circ$ $(e=0.5)$ for position and orientation data, respectively, and with scale and shape parameter optimization restricted to within $[0.7, 1.3]$ and $\pm1$ SD, respectively. We assign confidence to the registrations based on the tests explained in Sec.~\ref{sec:method} and validated in Sec.~\ref{sec:exp_sim}. 

All registrations run with $0$ modes terminated at the maximum iteration threshold of $100$, while those run using modes converged at an average $10.36$ iterations in $26.03$ seconds. Fig.~\ref{fig:clinic} shows registrations using increasing modes from left to right for each sequence plotted against $\mathrm{E}_p$ (middle) and $\mathrm{E}_o$ (right). All deformable registration results pass the $\mathrm{E}_p$ test as they fall below the 
$p=0.95$ threshold (Fig.~\ref{fig:clinic}, middle) using the chi-square inverse test. However, several of these fail the $\mathrm{E}_o$ test (Fig.~\ref{fig:clinic}, right). Deformable registrations on sequence 01 using $50$ modes 
and on sequence 04 for all except $30$ modes pass this test with \textbf{low confidence}. Using $30$ modes, the registration on sequence 04 passes \textbf{somewhat confidently}. The rigid registration on sequence 04 (the only rigid registration to pass both $\mathrm{E}_p$ and $\mathrm{E}_o$) and all deformable registrations on sequence 05 pass this test \textbf{very confidently}. Although, the rigid registration on sequence 05 passes this test very confidently, $\mathrm{E}_p$ already labels it a failed registration. 
Successful registrations produced a mean residual error of $0.78$ ($\pm0.07$)\,mm. Visualizations of successful registrations also show accurate alignment (Fig~\ref{fig:clinic}, left).


%
\section{Conclusion}
\label{sec:results}
We show that GD-IMLOP is able to produce submillimeter registrations 
in both simulation and in-vivo experiments, and assign confidence to these registrations. Further, it can accurately predict the anatomy where video data is available. In the future, we hope to learn statistics from thousands of CTs to better cover the range of anatomical variations. 
Additional features like contours can also be used 
to further improve registration and to add an additional test to evaluate the success of the registration based on 
contour alignment. Using improved statistics and reconstructions from video along with confidence assignment, this approach can be extended for use 
in place of CTs during endoscopic procedures.

\section*{Acknowledgment}
\label{sec:acknow}
This work was funded by NIH R01-EB015530, NSF Graduate Research Fellowship Program, an Intuitive Surgical, Inc. fellowship, and JHU internal funds.
%
%
%
\bibliographystyle{IEEEtran}  
\bibliography{EndNav_woCT.bbl}

\end{document}